# Nonintegrability and quantum fluctuations in a quantum optical model

Nilakantha Meher[a] and S. Sivakumar[b]

Theoretical Studies Section, Materials Physics Division, Indira Gandhi Centre for Atomic Research

Kalpakkam- 603102 INDIA

Email: [a] nilakantha.meher6@gmail.com, [b] siva@igcar.gov.in

Integrability in quantum theory has been defined in more than one ways. Recently, Braak suggested a new definition that a quantum system is integrable if the number of parameters required to specify the eigenstates and the number degrees of freedom (both discrete and continuous) are equal. It is argued that the dependence of uncertainty product of suitable operators on the atom-field interaction strength is distinctly different for the integrable and nonintegrable cases. These studies indicate that uncertainty product is able to identify nonintegrable atom-field systems in the context of the new definition.

## Introduction

A classical dynamical system with $n$ degrees of freedom (DOF) is integrable, Liouvillean integrable to be precise, if there are equal number of suitable constants of motion (COM) that have vanishing Poisson bracket among themselves and with the Hamiltonian[1]. Otherwise, the system is nonintegrable. While this definition is based on a sound mathematical footing, the situation in quantum dynamics is not very clear, essentially arising from the difficulty in defining or identifying DOF in quantum theory[2]. One possibility is to define integrablity by the existence of sufficient number of observables which commute with the Hamiltonian and pair-wise commute among themselves. However, this is wrought with difficulties as it may not be possible to arrive at classical limits of some quantum systems as in the case of a single two-level atom interacting with a single mode of the electromagnetic field. The former is a discrete DOF (Hilbert space of finite dimension) and the later is a continuous DOF (infinite dimensional ). While the continuous DOF has a proper classical limit, the two-level atom does not have a suitable classical limit.

According to a new definition introduced by Braak, *a system is integrable if the number of parameters required to specify the eigenstates of the Hamiltonian is equal to the sum of the number of discrete DOF and continuous DOF*[2]. This definition does not involve the existence of constants of motion, though all such cases are covered by this definition. In this new definition of integrability, some of the nonintegrable systems based on the Liouvillean definition become integrable. A simple example of such a system is the Rabi model describing the interaction between a two-level atom and a single mode of the electromagnetic field described by the Hamiltonian[3];

$$H_R = \frac{1}{2}\hbar\omega_0\sigma_z + \hbar\omega a^\dagger a + g\sigma_x(a + a^\dagger). \tag{1}$$

Here, $\sigma_{x,z}$ are Pauli matrices, $\omega_0$ is the atomic transition frequency, $a$ ($a^\dagger$) denote the annihilation (creation) operators of field with frequency $\omega$ and $g$ is the atom-field coupling constant. This Hamiltonian has only one COM, namely, $H_R$ itself. Since there are two DOF, the field and the two-level atom, the Hamiltonian is nonintegrable in the sense of Liouville. However, exploiting the parity symmetry in $H_R$, the system has been shown to be integrable[2]. Another interesting case is the rotating wave approximation of $H_R$, yielding the well known Jaynes-Cummings model[4,5]: $H_{JC} = \frac{1}{2}\hbar\omega_0\sigma_z + \hbar\omega a^\dagger a + g(\sigma_+ a + \sigma_- a^\dagger)$. This Hamiltonian has two COM, the Hamiltonian $H_{JC}$ and the operator for the number of excitations $N = a^\dagger a + |e\rangle\langle e|$. Existence of these two COM renders the Hamiltonian integrable. The eigenstates are labelled by two parameters, an integer $n$ corresponding to the number of excitations and the total energy. Also, both the Hamiltonians $H_R$ and $H_{JC}$ exhibit level-crossings of the eigenvalues as the interaction strength $g$ is varied, which is an indication that the models are integrable[2]. Level-crossing refers to the phenomenon wherein the eigenvalues depend on the interaction strength $g$ in such a way that the eigenvalues corresponding to two different eigenstates become degenerate at a specific value of $g$ and reverse their order for other values of $g$.

**Nonintegrable Model**

An interesting modification to $H_R$ making it nonintegrable is to break the parity symmetry by adding $\varepsilon\sigma_x$ to $H_R$ defined in Eq, (1), giving a new Hamiltonian $H_\varepsilon$. Within the scope of the Braak's definition of integrability, this model is integrable only when $\varepsilon$ is an integral multiple of $\omega/2$. This is also borne out by the existence of level crossings as shown in Fig.1. This criterion is sufficient for nonintegrability. We assume resonance, $\omega_0 = \omega$. For the results presented here, it is assumed that $\hbar = 1$ and $\omega = 1$. In Fig. 1, the absence and presence of level-crossing indicates respectively the nonintegrability and integrability of the Hamiltonian $H_\varepsilon$.

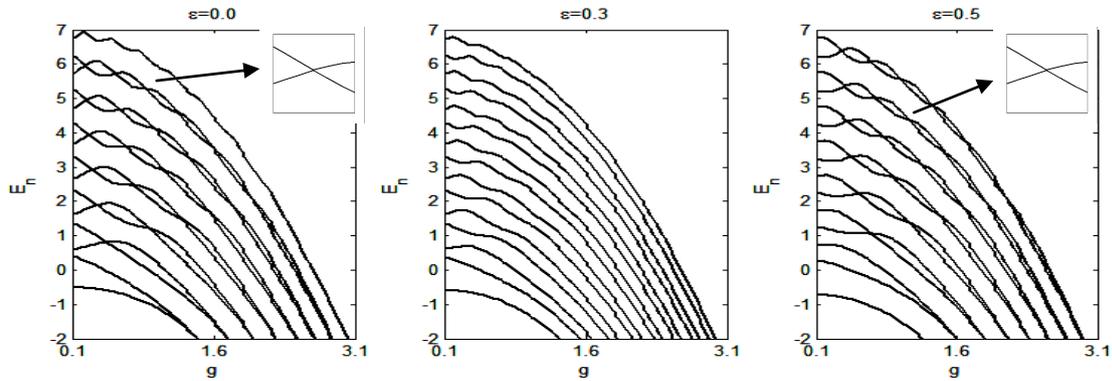

*Fig.1.Energy level ($E_n$) as a function of $g$ for different $\varepsilon$. Level crossing occurs if $\varepsilon =0$ and $0.5$ indicating integrability. No level crossing if $\varepsilon = 0.3$, indicating nonintegrability. Inset shows larger view of level crossing.*

A pertinent question in this context is to know those features that distinguish a nonintegrable atom-field system from an integrable one. One answer to this query appears to be that uncertainty product of a pair of suitably defined operators show markedly different characteristics. Since the system is nonintegrable, it is formidable to construct an analytical solution. Therefore, extensive numerical experimentations have been carried out and the results are presented here which support the claim stated above. Nonintegrability being a feature of the

Hamiltonian, it is natural to expect that the eigenstates carry signatures revealing this feature. To explore this, we define two self-adjoint operators of the two-level atom, $\tilde{\sigma}_x = \frac{1}{\sqrt{2}}(\sigma_+ + \sigma_-)$, $\tilde{\sigma}_y = \frac{i}{\sqrt{2}}(\sigma_+ - \sigma_-)$, where $\sigma_+$ ($\sigma_-$) is the atomic raising (lowering) operator. The commutation relation $[\tilde{\sigma}_x, \tilde{\sigma}_y] = -i\sigma_z$ implies that the value of the product of uncertainties lies between 0 and 1/2. The uncertainty relation of above operators is $\Delta = \Delta\tilde{\sigma}_x \Delta\tilde{\sigma}_y = \sqrt{\langle\tilde{\sigma}_x^2\rangle - \langle\tilde{\sigma}_x\rangle^2}\sqrt{\langle\tilde{\sigma}_y^2\rangle - \langle\tilde{\sigma}_y\rangle^2}$ .where $\langle\bullet\rangle$ is expectation value in any eigenstate. In Fig. 2, the uncertainty product $\Delta$ is plotted as a function of the atom-field interaction strength for different values of $\varepsilon$: $\varepsilon = 0$, 0.5 and 1.0 corresponding to the integrable cases and a few other values of $\varepsilon$ corresponding to nonintegrable cases. It is seen that as the parameter $g$ increases, the uncertainty product attains its maximum allowed value of ½ for the integrable cases. On the other hand, for the nonintegrable cases the uncertainty product falls below the limit of ½. In order to establish that the uncertainty product is very sensitive to the nature of the integrable and nonintegrable cases, the plots corresponding to values of $\varepsilon$ very close to integrable cases have been chosen.

For instance, in the second row in Fig. 2, the sudden change in the nature of uncertainty product as $\varepsilon$ assumes values 0.49 (nonintegrable) 0.5 (integrable) and 0.51 (nonintegrable) respectively are shown. In order to bring out the creatures more clearly, the probability distribution of the uncertainty products in different eigenstates are shown Fig. 3 corresponding to the respective figures in Fig. 2. The sharply peaked probability distribution indicates integrability.

**Summary**

Identification of nonintegrability in an interacting atom-field system is possible by the concentration of uncertainty product near a particular value as the atom-field interaction strength is increased. This feature seems to be related closely to the nonintegrability, also supported the absence of level crossings. This feature has been found to be able to identify nonintegrability in many other models that have been studied, though the results are not presented here. . In essence, suitable uncertainty product is able to identify nonintegrability, which is often difficult to establish analytically or numerically. Nevertheless, our analyses raise some important questions for which answers are to be found: Is it possible to arrive at the existence of this feature using only the definition of nonintegrability used here? Given a Hamiltonian, how to identify the correct observables whose uncertainty product will concentrate as the interaction strength is increased? How to extend this idea if the number of atoms is larger?

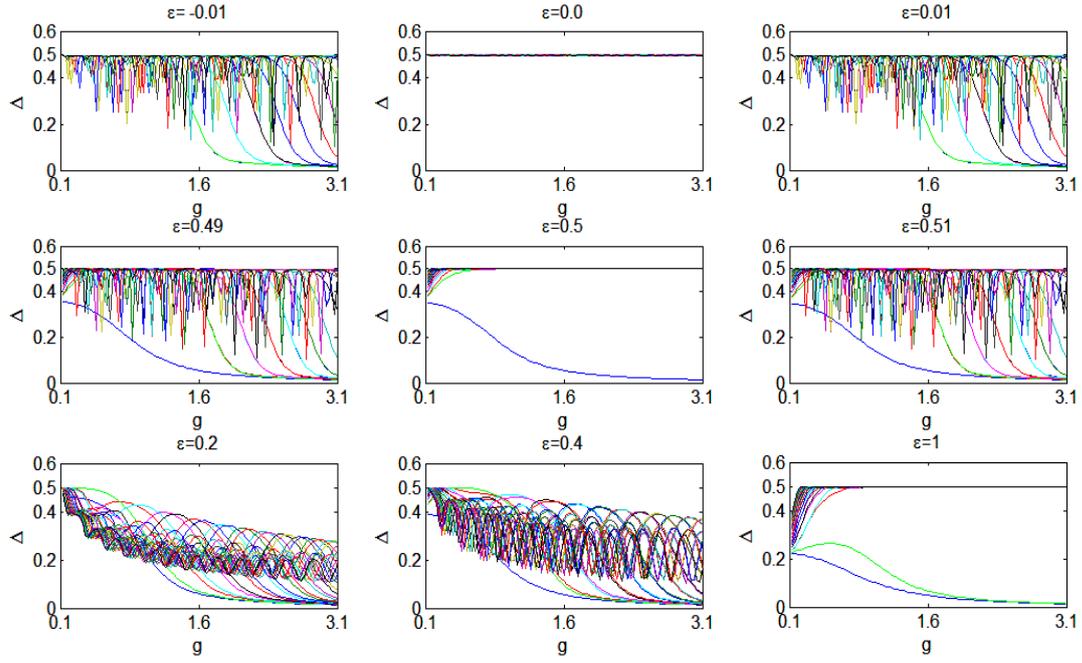

Fig.2.Uncertainty product ( Δ ) as a function of the atom-field coupling constant g. Different plots correspond to different values of ε: integrable cases: ε=0, 0.5 and 1.0, nonintegrable cases: ε = -0.01,0.01, 0.49,0.51,0.2 and 0.4. In any plot, the uncertainty is plotted for the eingenstates corresponding to the first fifty eigenvalues.

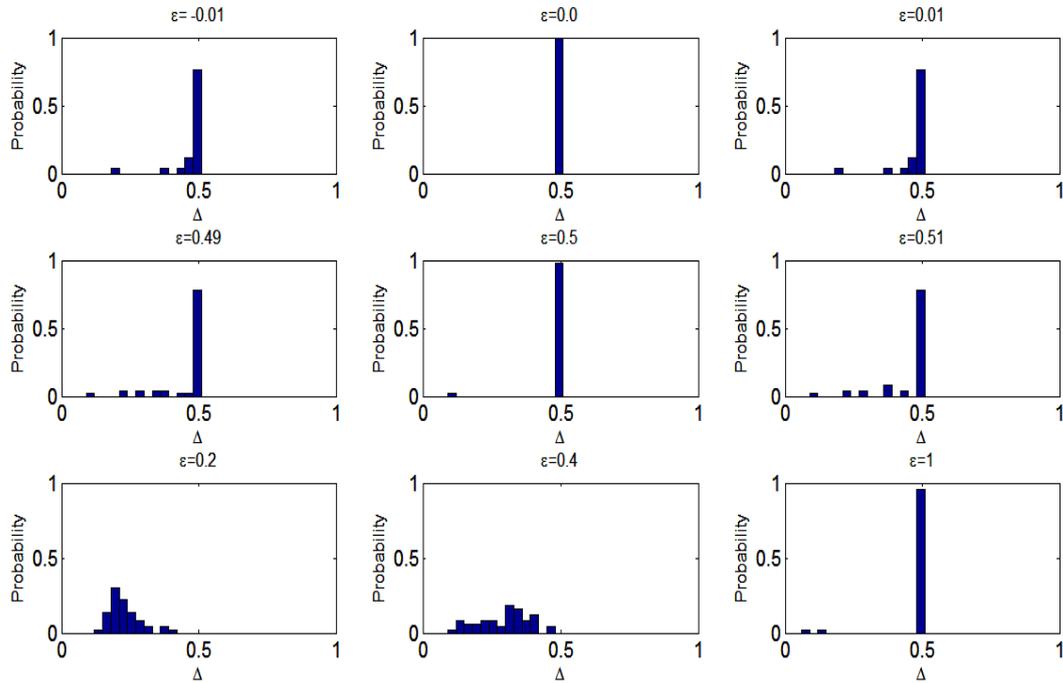

Fig.3. Probability distribution of the uncertainty product for all the eigenstates for a particular value of g, chosen to be 1.2 here. Any higher value of g yields the same results.